\documentclass[aps,prl,twocolumn,groupedaddress,showpacs,superscriptaddress]{revtex4-1}
\usepackage{graphicx,amsmath}
\usepackage{amssymb}

\begin{document}
\title{Accelerated Continuous time quantum Monte Carlo method with Machine Learning}
\author{Taegeun Song}
\affiliation{Department of Physics, Pohang University of Science and Technology, Pohang 37673, Republic of Korea}
\author{Hunpyo Lee}
\affiliation{School of General Studies, Kangwon National University, Samcheok, 25913, Republic of Korea}
\email{Email: hplee@kangwon.ac.kr}
\date{\today}

\begin{abstract}
An acceleration of continuous time quantum Monte Carlo (CTQMC) methods is a potentially interesting branch of work as they are 
matchless as impurity solvers of a density functional theory in combination with a dynamical mean field theory approach for the 
description of electronic structures of strongly correlated materials. The inversion of the $k \times k$ matrix given by the diagram 
expansion order $k$ in the CTQMC update and the multiplication of the $k \times k$ matrix and the non-interacting Green's 
function to measure the impurity Green's function are computationally time-consuming. Here, we propose the CTQMC method in 
combination with a machine learning technique, which would eliminate the need for multiplication of the matrix with 
the non-interacting Green's function. This method predicts the accurate impurity Green's function and double occupancy at low 
temperature, and also considers the physical properties of high Matsubara frequency in a much shorter computational time than the 
conventional CTQMC method.

\end{abstract}

\maketitle

{\it Introduction.-} Gaining an understanding of unconventional behaviors observed in strongly correlated electronic materials is an 
interesting and challenging research subject in theoretical condensed matter physics and material science communities, 
because exact diagonalization of a Hamiltonian with an electronic correlation $U$ requires extremely huge numerical work with 
increasing lattice size~\cite{Imada1998}. Dynamical mean field theory (DMFT) approximation is a useful tool in exploring such a 
Hamiltonian, although it ignores the spatial correlations to avoid excessive numerical burden~\cite{Georges1996}. DMFT captures 
unconventional behaviors such as a Mott insulator and a non-Fermi-liquid state beyond the mean field approximations, as well as 
supplies the reliable spectral functions in combination with the first-principles density functional theory method, which can be 
directly compared with angle-resolved photoemission spectroscopy experimental ones of real correlated materials such as iron-based 
superconductors and transition metal
oxides~\cite{Kotliar2006,Han2018,Kim2018,Brito2016,Medici2014,Karolak2015,Sim2018,Ruff2013,Lee2012}.

The most challenging part of DMFT approximation is to solve a quantum impurity problem that describes how the electrons on an 
impurity site interact with ones in a thermodynamic bath. In this sense, developments of the exact continuous-time quantum Monte 
Carlo (CTQMC) approaches for the quantum impurity problem have made remarkable progress in DMFT as well as in strongly correlated 
electronic material communities~\cite{Rubtsov2005,Werner2006,Gull2011}. The unconventional natures of many strongly correlated 
materials could be explained by the density functional theory plus DMFT (DFT+DMFT) tools in combination with CTQMC 
methods~\cite{Kotliar2006,Han2018,Kim2018,Brito2016,Medici2014,Karolak2015,Sim2018,Ruff2013,Lee2012}.  On the other hand, the CTQMC 
tools still retain heavy computational works in the multi-orbital systems with low temperature $T$ and high Matsubara 
frequencies~\cite{Rubtsov2005,Werner2006,Gull2011}. Therefore, accelerating the CTQMC algorithm for those systems would be highly 
beneficial.

Computationally, the CTQMC methods have two time-consuming parts: (i) One is the inversion of the $k \times k$ matrix given by the 
diagram expansion order $k$ in the CTQMC update. (ii) Another one is the multiplication of the $k \times k$ 
matrix and the non-interacting Green's function $G_{\sigma}^0 (i\omega_n)$ in order to measure the impurity Green's function 
$G_{\sigma} (i\omega_n)$, where $\sigma$ and $\omega_n$ are the spin index and the Matsubara frequency, respectively. Computational 
time in the second one increases proportionally with increasing values of $\omega_n$ and $k$.

\begin{figure}
\includegraphics[width=1.0\columnwidth]{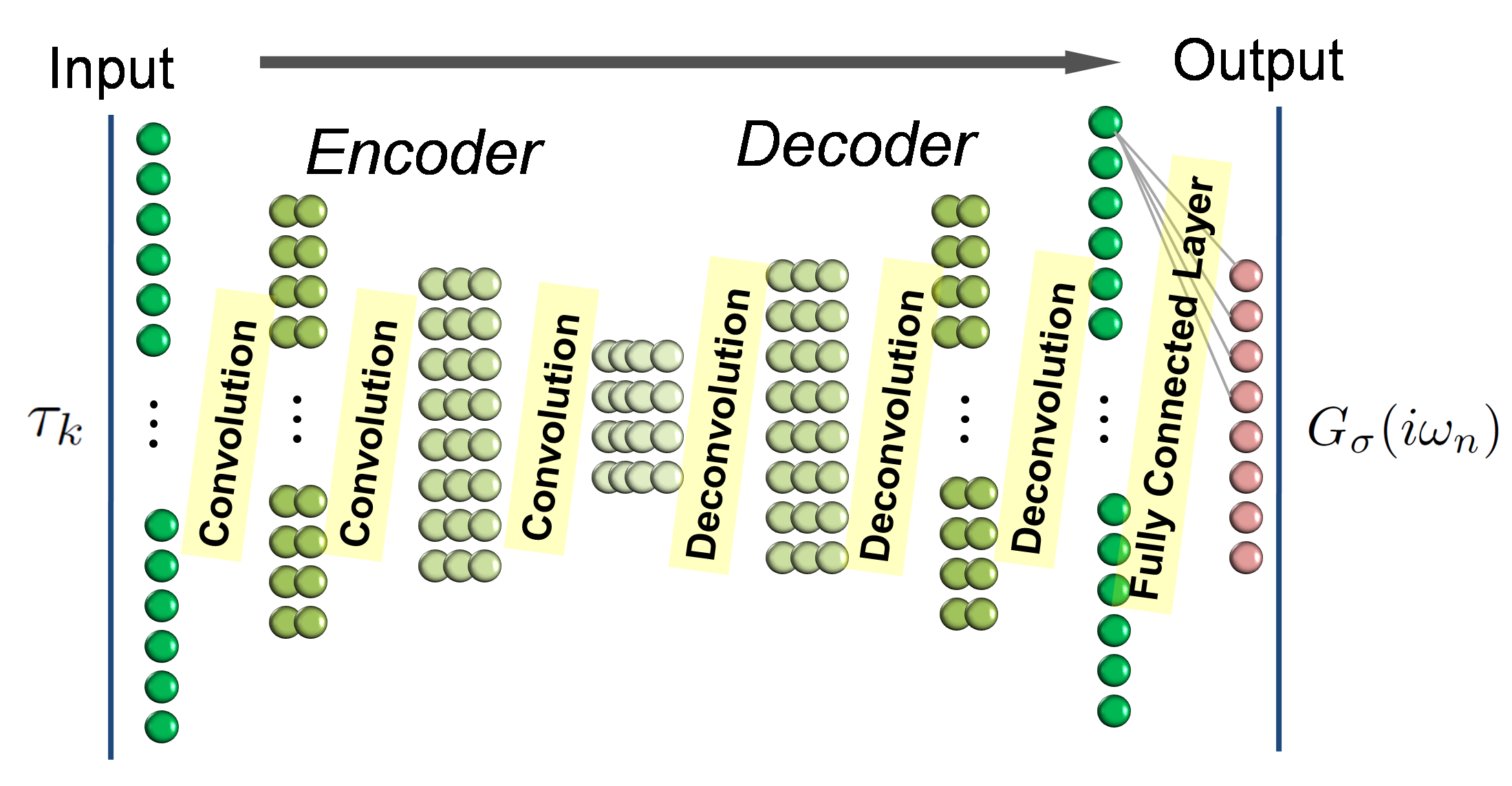}
\caption{\label{Fig1} Schematic architecture of one-dimension convolutional autoencoder. The input imaginary time $\tau_k$, which 
is obtained from a continuous time quantum Monte Carlo (CTQMC) approach, is an array of a $k$ size, where $k$ is a diagram 
expansion order. The deep layered structure with three convolution and deconvolution layers is employed for a machine learning (ML) 
process. After the final deconvolution layer, we add one fully connected layer to match the desired number of output nodes. The 
impurity Green's functions $G_{\sigma} (i\omega_n)$ as a function of Matsubara frequency $\omega_n$ are output datasets.}
\label{fig1}
\end{figure}

Recently, modern machine learning (ML) approaches, which conjecture the results from massive data, have been extensively 
employed in various fields including that of theoretical condensed matter 
physics~\cite{Carrasquilla2017,Arsenault2014,Carleo12017,Yoon2018,Huang2017,Liu2017,Otsuki2017,Amin2018}. Here, we propose the CTQMC 
method in combination with the ML (CTQMC+ML) technique, which eliminates the time-consuming step of of the matrix. The CTQMC+ML 
algorithm is as follows. The raw datasets of $G_{\sigma} (i\omega_n)$ obtained from the data of the imaginary time $\tau_k$ with an 
array of $k$ order are constructed by the CTQMC simulation. $\tau_k$ and $G_{\sigma} (i\omega_n)$ are employed for input data and 
output data, respectively. The ML kernel is constructed through a one dimension convolutional autoencoder (CAE) and a stochastic 
gradient descent based optimizer implemented in Tensorflow-gpu~\cite{tsflow, LeCun2015,Krizhevsky2012,Bengio2014}.
After finishing the training of ML, $G_{\sigma} (i\omega_n)$ is directly predicted by the ML process, where the multiplication of 
the $k \times k$ matrix and the non-interacting Green's function $G_{\sigma}^0 (i\omega_n)$ with the number of $\omega_n$ disappears 
in the calculation. Finally, we find that our ML results show accurate $G_{\sigma} (i\omega_n)$ values with high Matsubara 
frequency and double occupancy in a much shorter computational time than the CTQMC approach.

{\it Continuous time quantum Monte Carlo method in combination with Machine learning.-}  The Hamiltonian $H$ in the DMFT 
approximation includes $H_{\text {Local}}$, $H_{\text {Bath}}$, and $H_{\text {Hyb}}$ for an impurity part, a thermodynamic bath 
part, and a coupling part between the impurity and the thermodynamic bath, respectively. The CTQMC approaches have two versions, one 
where the partition functions are expanded on the interaction or hybridization. Although the expansion quantities differ rather 
significantly, the simulation procedures of both algorithms are quite similar. Therefore, we will explain the interaction-expansion 
CTQMC method in combination with ML, and discuss the results of $G_{\sigma}(i\omega_n)$ and double occupancy.

Here, we discuss the formalism of the interaction-expansion CTQMC tool briefly~\cite{Rubtsov2005}. The partition function of an 
impurity problem is written as
\begin{equation}
Z = \int D[d^{\dagger},d] \exp(-S_0 - S_U),
\end{equation}
where $S_0$ and $S_U$ are the non-interacting part including the hybridization and the electronic interaction part given as
$S_U = U\int_0^{T^{-1}} d\tau (n_{\uparrow} (\tau) - \frac{1}{2}) (n_{\downarrow} (\tau) - \frac{1}{2})$,
respectively. Here, for simplicity of the notation to remove the Fermionic sign problem in the cases of the repulsive interaction 
$(U>0)$ we introduce $n_{\uparrow}^{'}n_{\downarrow}^{'}$ operator given as 
$n_{\uparrow}^{'}n_{\downarrow}^{'}=U(n_{\uparrow} - \frac{1}{2}) (n_{\downarrow} - \frac{1}{2})=\frac{U}{2}n_{\uparrow}
(n_{\downarrow}-1)+\frac{U}{2}(n_{\uparrow}-1)
n_{\downarrow}$. The expansion of the partition function in the powers of the modified interaction part is given as
\begin{equation}
\frac{Z}{Z_0} = \sum_{C_k} (-U)^k <n_{1\uparrow}^{'} (\tau_1) n_{1\downarrow}^{'}(\tau_1) \dots n_{k\uparrow}^{'}
(\tau_k)n_{k\downarrow}^{'}(\tau_k)>_0,  
\end{equation}
where $\sum_{C_k} = \sum_{k=0}^{\infty} \int_0^{T^{-1}} d\tau_1  \dots \int_0^{\tau_{k-1}} d\tau_k$ and 
$Z_0=\int D[d^{\dagger},d]e^{-S_0}$ is the single-site case. The CTQMC update is performed on the diagram expansion order $k$ and 
the continuous time $\tau$ occurred between $0$ and $T^{-1}$. $<n_{1\uparrow}^{'}(\tau_1)n_{1\downarrow}^{'}(\tau_1) \dots 
n_{k\uparrow}^{'} (\tau_k)n_{k\downarrow}^{'}(\tau_k)>_0$ is a determinant of a matrix obtained from the Wick's theorem, and the 
matrix can be separated into spin up matrix $D_{\uparrow}$ and spin down matrix $D_{\downarrow}$. The impurity Green's function 
$G_{\sigma} (i\omega_n)$ can be expressed as 
\begin{equation}
G_{\sigma}(\tau)  = \frac{<Tc_{\sigma}^{\dagger}(\tau)c_{\sigma}(0)c_{\sigma}^{\dagger}(\tau_1)c_{\sigma}(\tau_1)\dots 
c_{\sigma}^{\dagger}(\tau_k)c_{\sigma}(\tau_k)>_0}{\det D_{\sigma}}. 
\end{equation}
In the end, via the Fourier transformation and the Wick's theorem, $G_{\sigma} (i\omega_n)$ as a function of the Matsubara frequency 
is computed with
\begin{equation}\label{IG}
G_{\sigma} (i\omega_n) = G_{\sigma}^0(i\omega_n) - G_{\sigma}^0(i\omega_n) [\sum_{i,j} M_{i,j,\sigma} e^{i\omega_n 
(\tau_i - \tau_j)}] G_{\sigma}^0(i\omega_n), 
\end{equation}
where the size $k \times k$ matrix $M_{i,j,\sigma}$ is the inversion of $D_{\sigma}/T$. The average matrix size $<k>$ is 
proportional to $T^{-1}$. Note that most computational times in the CTQMC algorithm are used to calculate the inversion of 
$k \times k$ matrix $D_{\sigma}/T$ and the multiplication of the matrix in Eq.~(\ref{IG}).

\begin{figure}
\includegraphics[width=1.0\columnwidth]{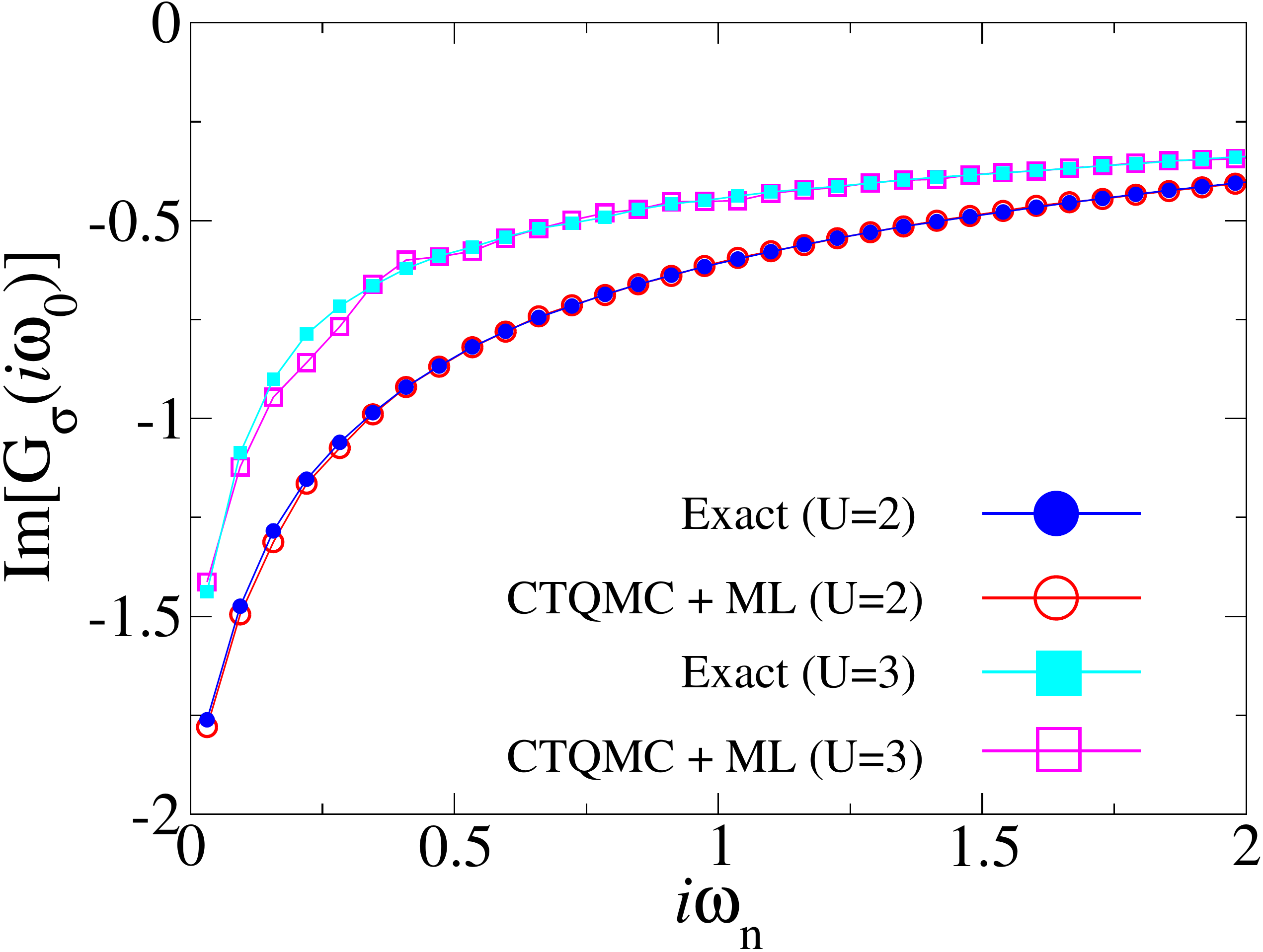}
\caption {\label{Fig2} (Color online) Imaginary part of the impurity Green's function $\text{Im} [G(i\omega_n)]$ as a function of 
$\omega_n$ at half-filling for $U=2.0$ and $3.0$. 'Exact' means that the exact results are obtained from Eq.~(\ref{IG}). 'CTQMC+ML' 
is the continuous time quantum Monte Carlo in combination with the machine learning. The trainings of ML for $U=2.0$ and $3.0$ are 
independently performed to obtain each $\text{Im} [G_{\sigma} (i\omega_n)]$. We use the conventional CTQMC measurements of 
Eq.~(\ref{IG}) with $6\times10^5$ and $8 \times 10^4$ different combinations of dataset for exact and CTQMC+ML results, 
respectively. The bandwidth $W$ and the temperature $T$ are set to $W = 2.0$ and $T = 0.01$, respectively. The real parts of the 
impurity Green's functions are set to $0$ at half-filling.}
\end{figure}

Here, we employ the ML technique to substitute for the calculation of Eq.~(\ref{IG}). To construct the ML kernel, we adapt the CAE,
which is customized as a one dimensional input layer with values of the $k$ size imaginary time $\tau_k$ and deep layered structure 
with three convolution and deconvolution layers. Figure~\ref{Fig1} shows the schematic CAE architecture from input data $\tau_k$ to 
output data $G_{\sigma} (i\omega_n)$ via deep layered structure\cite{Bengio2014,Kingma2014}. During the convolution process, we 
apply max pooling and keep the probability larger than $0.7$ with a filter of $1 \times 5 \times m$, where $m$ denotes a feature map 
$1-16-32-64-32-16-1$ of ongoing progress. After the deconvolution layer, we add one fully connected layer to match the desired 
number of output nodes. The softmax activation and stochastic gradient descent based optimizer are applied for the one-dimensional 
CAE training implemented with the TensorFlow-gpu version\cite{tsflow}. We prepare $8\times 10^4$ different combinations of $\tau_k$ 
as the input dataset and use batch learning with a few thousand parallel procedures. In all cases, the training times of ML process 
take approximately one hour with NVIDIA GeForce-1080Ti. The results of the CTQMC+ML approach are compared with those of conventional 
CTQMC measurements of Eq.~(\ref{IG}) with $6\times10^5$.

{\it Results.-} We consider the single impurity system on the Bethe lattice with the semi-circular density of states. We set the 
bandwidth to $W=2.0$ and the temperature $T=0.01$ in all calculations. In these cases the non-interacting Green's function 
$G_{\sigma}^0 (i\omega_n)$ with the chemical potential $\mu$ is given as 
\begin{equation}\label{NG}
G_{\sigma}^0(i\omega_n) = \frac{2}{i\omega_n + \sqrt{\omega_n^2 + 1}+2\mu}.
\end{equation}

We study the cases of half-filling with $\mu=0$ in Eq.~(\ref{NG}). Figure~\ref{Fig2} shows the imaginary part of the impurity 
Green's function $\text{Im} [G_{\sigma} (i\omega_n)]$ obtained from exact Eq.~(\ref{IG}) and the CTQMC+ML approach with $8 \times 
10^4$ different trained datasets for $U=2.0$ and $3.0$. The trainings of the ML for $U=2.0$ and $3.0$ are independently performed to 
obtain each $\text{Im} [G_{\sigma} (i\omega_n)]$. We approximately spend one hour for the training of the ML process. (i) We find 
that the computational time in the measurement step of $G_{\sigma}(i\omega_n)$ of the CTQMC+ML method is almost independent, 
regardless of the numbers of $k$ order and $\omega_n$. It means that the computational time of the CTQMC+ML method is much less than 
one of the conventional CTQMC approaches with proportional increase, as the numbers of $\omega_n$ and $k$ increase. (ii) Although we 
do not show the CTQMC+ML results obtained from $2 \times 10^4$ different trained datasets in Figure~\ref{Fig2}, $\text{Im} 
[G_{\sigma} (i\omega_n)]$ of both methods for $U=2.0$ are already in good agreement as results show in Figure~\ref{Fig2}, while they 
are quite different for $U=3.0$. This means that the numbers of the trained datasets affect the accuracy of the CTQMC+ML results 
more strongly than other parameters such as the batch size, the training time, and the filter length.

\begin{figure}
\includegraphics[width=1.0\columnwidth]{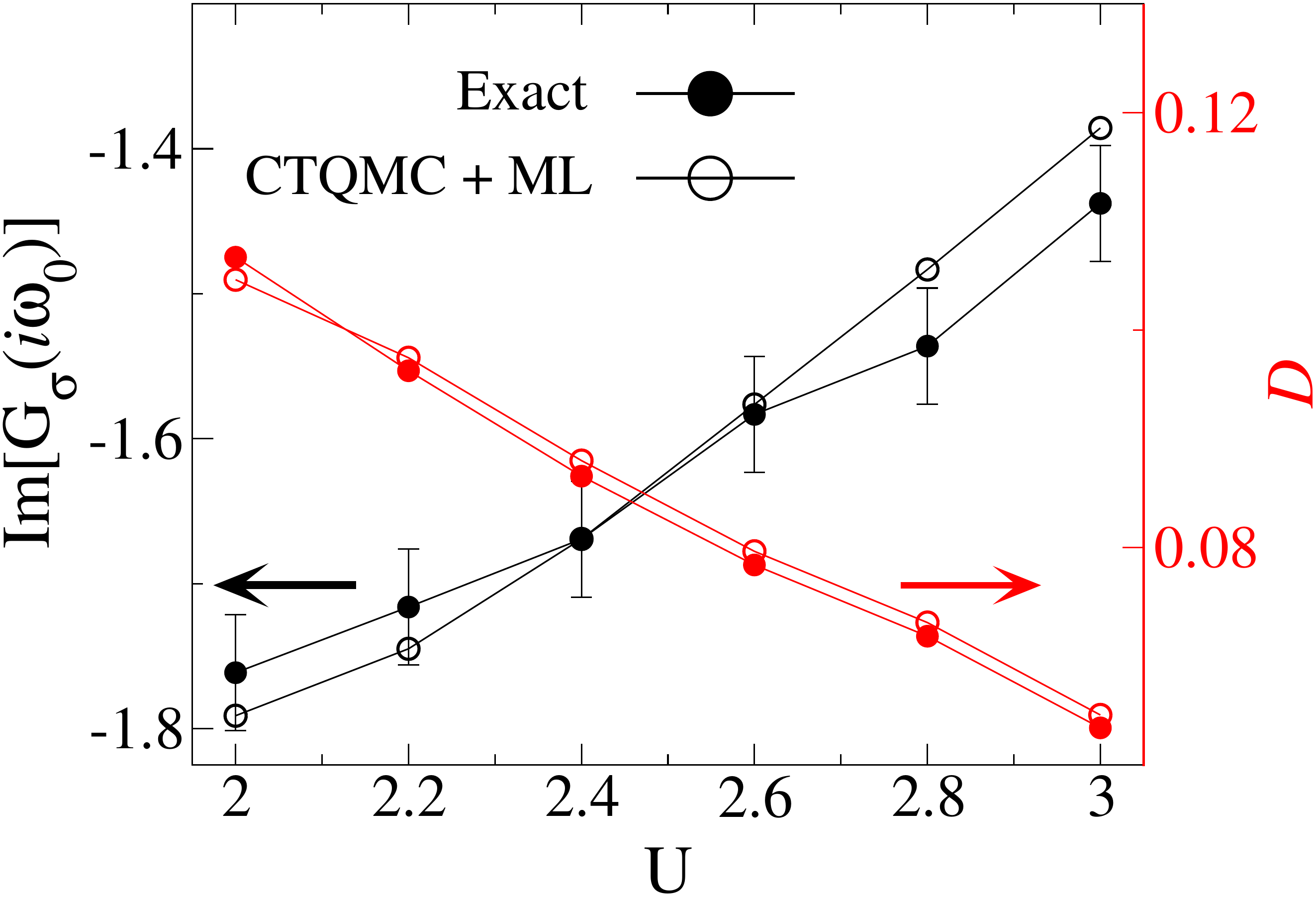}
\caption {\label{Fig3} (Color online) (Left) Imaginary part of the impurity Green's function $\text{Im} [G_{\sigma}(i\omega_0)]$ at 
the first Matsubara frequency $\omega_0$ and (Right) double occupancy $D$ as a function of $U$. The ML kernel is constructed on 
input and output data of both $U=2.0$ and $3.0$, unlike the case with independent ML trainings for each $U=2.0$ and $3.0$ in 
Figure~\ref{Fig2}. $\text{Im} [G_{\sigma} (i\omega_0)]$ and $D$ between $U=2.0$ and $3.0$ are predicted with the CTQMC+ML process.}
\end{figure}

Next, we perform the training of the ML with input and output datasets of both $U=2.0$ and $3.0$, and predict the double occupancy 
$D$ and $\text{Im} [G_{\sigma} (i\omega_n)]$ between $U=2.0$ and $3.0$. $D$ is given as 
\begin{equation}\label{NG1}
D= \frac{2<k>-UT^{-1}}{2UT^{-1}},
\end{equation}
where $<k>$ is an average diagram expansion order. Although we already know $<k>$ in Eq.~(\ref{NG1}) while performing the 
CTQMC update, we estimate $<k>$ from ML approach again to compute $D$. The results of the imaginary part of impurity Green's 
function $\text{Im} [G_{\sigma} (i\omega_0)]$ at the first Matsubara frequency $\omega_0$ and the $D$ as a function of $U$ 
are shown in Figure~\ref{Fig3}. $D$ in all $U$ and $\text{Im} [G_{\sigma} (i\omega_0)]$ in the weak interaction 
are in a very good agreement, while $\text{Im} [G_{\sigma} (i\omega_0)]$ for $U=2.8$ and $3.0$ are barely in agreement within error 
the bar, owing to strong numerical fluctuations and lack of trained datasets. We believe that the results with strong numerical 
uncertainty will be improved with increasing the measurements in Eq.~(\ref{IG}) and the trained dataset in the CTQMC+ML method.

\begin{figure}
\includegraphics[width=1.0\columnwidth]{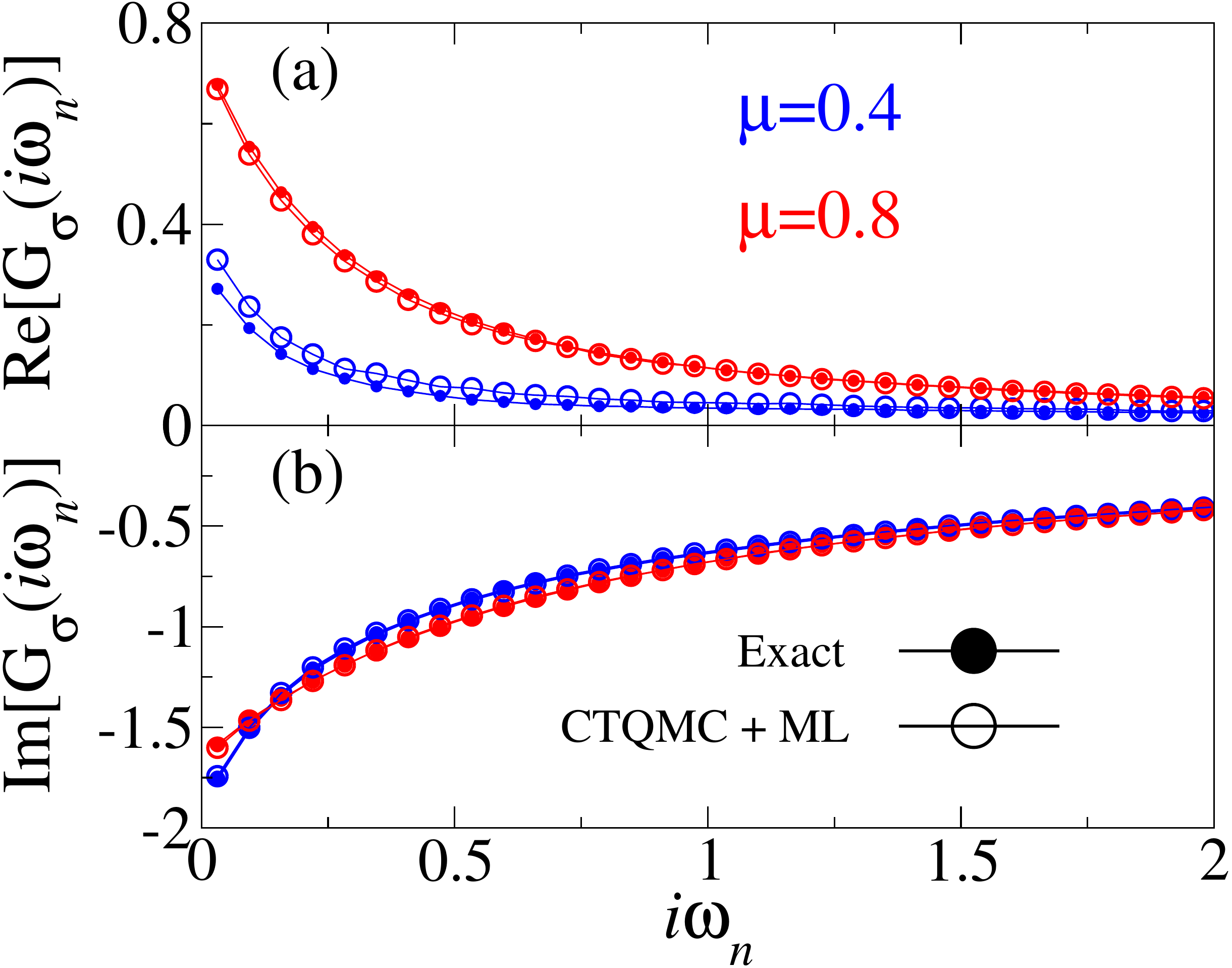}
\caption {\label{Fig4} (Color online) (a) and (b) show the real and the imaginary parts of $G_{\sigma} (i\omega_n)$ as a function of 
$\omega_n$, respectively for $U=2.0$. The chemical potentials $\mu$ employed in Eq. (\ref{NG}) for dopings are $0.4$ and $0.8$.}
\end{figure}

Finally, we confirm the cases of away half-filling for $U=2.0$, where real and imaginary parts of $G_{\sigma} (i\omega_n)$ as a 
function of $\omega_n$ are shown in Figure~\ref{Fig4} (a) and (b), respectively. We employ $\mu=0.4$ and $0.8$ as dopants. The 
results obtained from Eq.~(\ref{IG}) and CTQMC+ML approach are in a very good agreement, as in the cases of half-filling.

{\it Conclusion.-} The first-principles DFT+DMFT approaches have proven to be successful in describing physical properties of 
strongly correlated materials, while the numerically time-consuming impurity step is presented in the DMFT part of these 
calculations. Therefore, the acceleration of the exact continuous time quantum Monte Carlo (CTQMC) approach sustaining accuracy as 
the impurity solver of the DFT+DMFT method is a step forward for the multi-orbital systems with low $T$ and high $\omega_n$. Here, 
we propose the CTQMC method in combination with the ML technique based on a one dimensional CAE and a stochastic gradient descent 
based optimizer in Tensorflow-gpu. We find that our CTQMC+ML can accurately predict $G_{\sigma}(i\omega_n)$ and $D$, but also 
considers the physical properties of high Matsubara frequency in much less computational time than one of the conventional CTQMC 
approaches.

We only confirm the properties of the two-point correlators like $G_{\sigma} (i\omega_n)$ in the CTQMC+ML approach. One future work 
is immediately raising. The dynamical vortex and dual fermion approximations, which can capture the nonlocal correlations beyond 
DMFT approximation, require the four-point correlators in the CTQMC part, where the computational times are proportional to square 
of the numbers of $k$ and $\omega_n$~\cite{Toschi2007,Rubtsov2008,Rohringer2018}. We think that our CTQMC+ML idea will reduce the 
computational time of the four-point correlators more dramatically than one of the two-point correlators.   

\section{Acknowledgements}

This work was supported by the Kangwon National University and Ministry of Science through NRF-2018R1D1A1B07048139 (Hunpyo Lee), and 
the Ministry of Science, ICT and Future Planning through NRF-2017R1D1A1B03034600 (Taegeun Song).

\end{document}